\documentclass[aps,prl,twocolumn,superscriptaddress,floatfix,letter]{revtex4-1}
\usepackage[bookmarks=true,colorlinks,linkcolor=OrangeRed,urlcolor=NavyBlue,citecolor=RoyalBlue]{hyperref}
\usepackage{epsfig,amsmath,amssymb,color,dsfont,upgreek,physics}
\usepackage{subfiles}
\usepackage{mathrsfs}
\usepackage{mathtools}
\usepackage{bbold}
\usepackage[makeroom]{cancel}
\usepackage[caption=false]{subfig}
\usepackage{mathrsfs}
\usepackage{graphicx}
\usepackage{subfig}
\usepackage[countmax]{subfloat}
\usepackage[english]{babel}
\usepackage[dvipsnames]{xcolor}

\usepackage{braket}

\definecolor{mypurple}{rgb}{0.49,0.18,0.56}
\definecolor{mygold}{rgb}{0.93,0.69,0.13}
\definecolor{mygreen}{rgb}{0,0.5,0}
\definecolor{myblue}{rgb}{0,0,0.75}
\definecolor{mymagenta}{cmyk}{0,1,0,0.12}
\definecolor{mygray}{rgb}{0.5,0.5,0.5}

\usepackage{umoline}

\newcommand{\mbeq}{\overset{!}{=}}
\begin{document}

\title{Stabilizing Disorder-Free Localization}
\author{Jad C.~Halimeh}
\email{jad.halimeh@physik.lmu.de}
\affiliation{INO-CNR BEC Center and Department of Physics, University of Trento, Via Sommarive 14, I-38123 Trento, Italy}
\author{Hongzheng Zhao}
\affiliation{Blackett Laboratory, Imperial College London, London SW7 2AZ, United Kingdom}
\affiliation{Max-Planck-Institut f\"ur Physik komplexer Systeme, N\"othnitzer Stra\ss e 38, 01187 Dresden, Germany}
\author{Philipp Hauke}
\affiliation{INO-CNR BEC Center and Department of Physics, University of Trento, Via Sommarive 14, I-38123 Trento, Italy}
\author{Johannes Knolle}
\affiliation{Department of Physics, Technische Universit\"at M\"unchen, James-Franck-Straße 1, D-85748 Garching, Germany}
\affiliation{Munich Center for Quantum Science and Technology (MCQST), Schellingstra\ss e 4, D-80799 M\"unchen, Germany}
\affiliation{Blackett Laboratory, Imperial College London, London SW7 2AZ, United Kingdom}

\begin{abstract}
Disorder-free localization is a paradigm of nonergodicity in translation-invariant quantum many-body systems hosting gauge symmetries. The quench dynamics starting from simple initial states, which correspond to extensive superpositions of gauge superselection sectors, exhibits many-body localization with the system dynamically inducing its own disorder. An open question concerns the stability of disorder-free localization in the presence of gauge-breaking errors, and whether processes due to the latter can be controllably suppressed. Here, we show that translation-invariant \textit{single-body gauge terms} induce a quantum Zeno effect that reliably protects disorder-free localization against errors up to times at least polynomial in the protection strength. Our experimentally feasible scheme not only shows that disorder-free localization can be reliably stabilized, but also opens promising prospects for its observation in quantum simulators.
\end{abstract}
\date{\today}
\maketitle

\textit{Introduction.--} Generic nonintegrable quantum many-body models are expected to thermalize based on the eigenstate thermalization hypothesis (ETH) \cite{Rigol_review,Deutsch_review}, while systems with quenched disorder may violate ETH and display many-body localization (disorder-MBL) \cite{Basko2006,Nandkishore_review} leading to the absence of thermalization. This binary picture of ergodic versus nonergodic behavior has turned out to be only the surface of a rich variety of physical phenomena.

Even disorder-free nonintegrable models may exhibit dynamical localization under certain circumstances. For example, in the presence of a constant electric field, a chain of spinless fermions interacting via nearest-neighbor repulsion will exhibit so-called Stark-MBL in its quench dynamics \cite{Schulz2019}, where localization arises even in the absence of quenched disorder. Alternatively, even without the spatial inhomogeneity due to the electric field, disorder-free localization (DFL) has emerged as another paradigm for absence of thermalization in (non)integrable translation-invariant and spatially homogeneous quantum many-body systems hosting gauge symmetries \cite{Smith2017,Brenes2018}. In models of DFL \cite{smith2017absence,Metavitsiadis2017,Smith2018,Russomanno2020,Papaefstathiou2020,karpov2021disorder,hart2021logarithmic,Zhu2021}, simple translation-invariant states correspond to massive superpositions of gauge-symmetry superselection sectors. Even though there is no quenched disorder, in a standard quench setup localization arises due to the dynamically induced disorder over the extensive number of superselection sectors. In short, for DFL the local constraints of the gauge symmetry take on the role of local conserved quantities, e.g., in the form of fixed background charges, which appear as a discrete disorder potential in a typical sector. Despite it being a superposition over superselection sectors, such an initial state can usually be prepared as a product state and is therefore easily implementable in experiments.

\begin{figure}[t!]
	\centering
	\includegraphics[width=.45\textwidth]{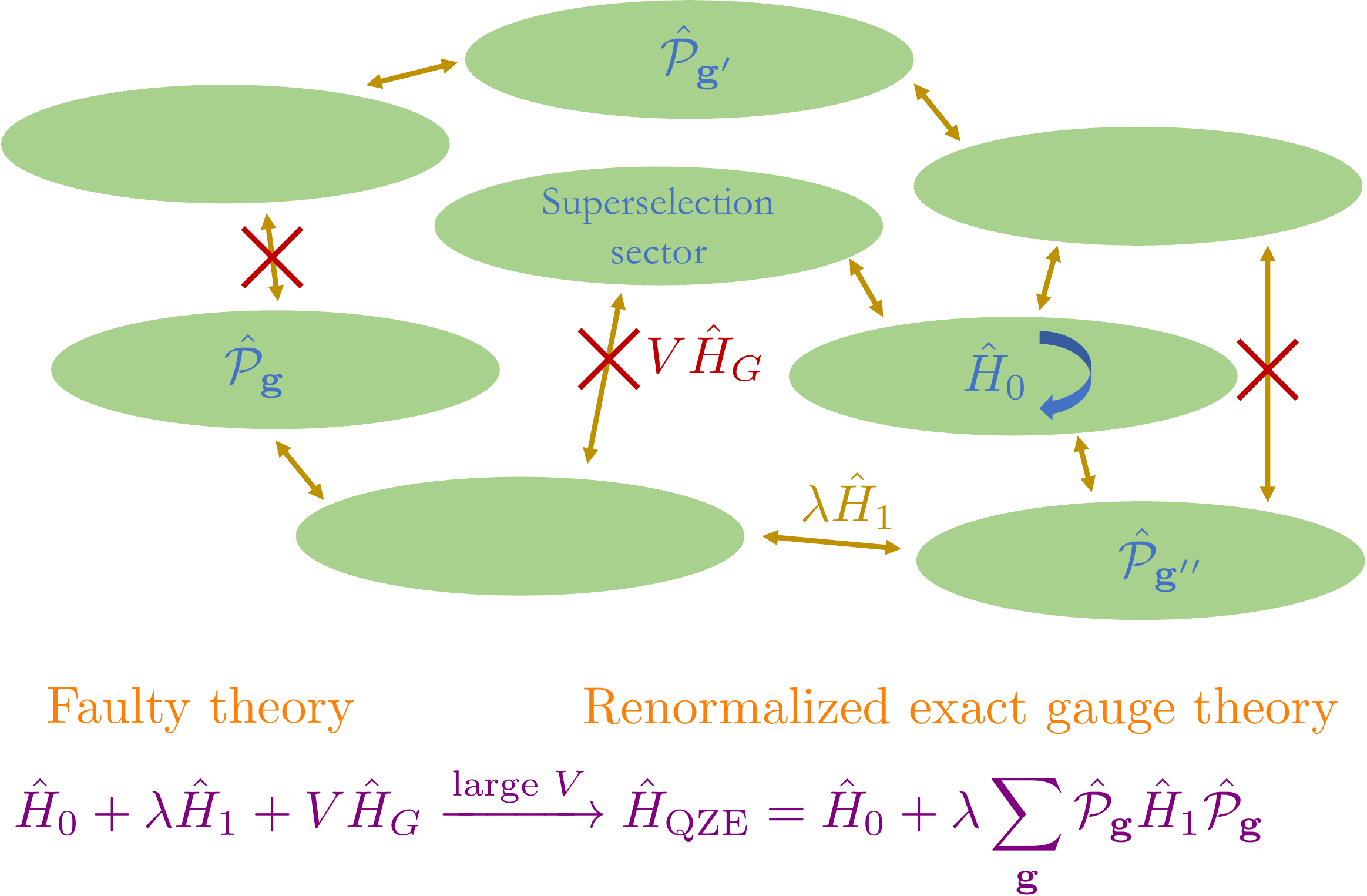}
	\caption{(Color online). Schematic of stabilization of disorder-free localization. The gauge theory $\hat{H}_0$ has a gauge-symmetry generator $\hat{G}_j$ with eigenvalues $g_j$ defining superselection sectors  $\mathbf{g}$ with projectors $\hat{\mathcal{P}}_\mathbf{g}$. Since $\big[\hat{H}_0,\hat{G}_j\big]=0,\,\forall j$, dynamics due to $\hat{H}_0$ does not couple different superselection sectors, leading to DFL when the initial state is a superposition over many sectors. Gauge-breaking errors $\lambda\hat{H}_1$ of strength $\lambda$ lead to transitions between the different superselection sectors, destroying DFL. Adding the translation-invariant single-body gauge protection $V\hat{H}_G=V\sum_j(-1)^j\hat{G}_j$ at strength $V$ induces quantum Zeno dynamics under a renormalized gauge theory $\hat{H}_\text{QZE}$, with the same gauge symmetry as $\hat{H}_0$, that suppresses inter-sector processes and stabilizes DFL up to timescales at least polynomial in $V$.}
	\label{fig:schematic} 
\end{figure}

Recently, there has been impressive effort in quantum synthetic matter (QSM) implementations of lattice gauge theories with dynamical matter and gauge fields \cite{Goerg2019,Schweizer2019,Mil2020,Yang2020,Zhou2021}. Gauge-breaking errors will always arise in such setups \cite{aidelsburger2021cold}, and these will ultimately destroy features such as DFL due to transitions between the different superselection sectors \cite{Smith2018}. It remains an outstanding question whether there exist means to actively control such errors and stabilize DFL. In this Letter, we employ the scheme of single-body gauge protection \cite{Halimeh2020e}, where a fully translation-invariant alternating sum of the gauge-symmetry generators suppresses transitions between different superselection sectors up to timescales at least polynomial in the protection strength based on the quantum Zeno effect (QZE) \cite{facchi2002quantum,facchi2004unification,facchi2009quantum,burgarth2019generalized}. In the process, DFL is restored by the dynamical emergence of a renormalized gauge theory with the same gauge symmetry as the ideal model, see Fig.~\ref{fig:schematic}.

\textit{Model.--} We consider the paradigmatic spin-$S$ $\mathrm{U}(1)$ quantum link model (QLM) \cite{Wiese_review,Chandrasekharan1997,Hauke2013,Yang2016,Kasper2017} in $(1+1)$ dimensions described by the Hamiltonian
\begin{align}\nonumber
\hat{H}_0=&\sum_{j=1}^L\bigg[\frac{J}{2\sqrt{S(S+1)}}\big(\hat{\sigma}^-_j\hat{s}^+_{j,j+1}\hat{\sigma}^-_{j+1}+\text{H.c.}\big)\\\label{eq:H0}
	&+\frac{\mu}{2}\hat{\sigma}^z_j+\frac{\kappa^2}{2}\big(\hat{s}^z_{j,j+1}\big)^2\bigg],
\end{align}
where $L$ is the total number of matter sites. This model mimics a lattice version of quantum electrodynamics (QED). The Pauli matrices $\hat{\sigma}_j^\pm$ denote the creation (annihilation) operators of the matter field on site $j$, where the matter occupation operator is $\hat{n}_j=(\hat{\sigma}_j^z+\hat{\mathds{1}}_j)/2$ and $\mu$ is the mass. In the above formulation, we have employed a Jordan--Wigner and particle-hole transformation~\cite{Hauke2013,Yang2016}. The gauge field at the link between sites $j$ and $j+1$ is represented by the spin-$S$ ladder operators $\hat{s}_{j,j+1}^\pm$, while the electric field is given by $\hat{s}^z_{j,j+1}$ with coupling strength $\kappa$. The last term in Eq.~\eqref{eq:H0} is irrelevant only for $S=1/2$ since then $(\hat{s}^z_{j,j+1})^2=\hat{\mathds{1}}_{j,j+1}$, leading to an inconsequential energetic constant. 

The $\mathrm{U}(1)$ gauge symmetry of $\hat{H}_0$ is generated by the operator
\begin{align}
\hat{G}_j=(-1)^j\big(\hat{n}_j+\hat{s}^z_{j-1,j}+\hat{s}^z_{j,j+1}\big),
\end{align}
where gauge invariance is embodied in the commutation relation $\big[\hat{H}_0,\hat{G}_j\big]=0,\,\forall j$. The eigenvalues $g_j$ of $\hat{G}_j$ are so-called background charges, and a configuration $\mathbf{g}=(g_1,g_2,\ldots,g_L)$ of them over the lattice defines a gauge superselection sector.

\begin{figure}[t!]
	\centering
	\includegraphics[width=.4\textwidth]{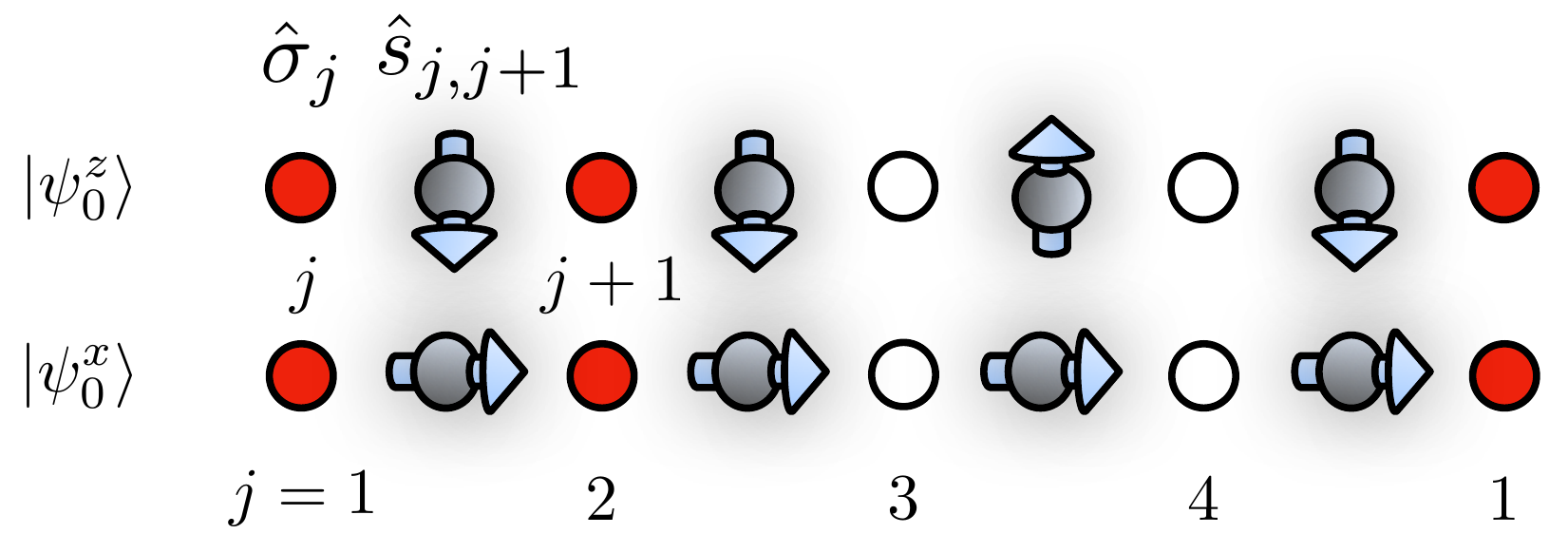}
	\caption{(Color online). Product states in which the system is initialized, with circles denoting matter fields (red-filled is occupied, otherwise empty) and arrows denoting the electric-field orientation: $\ket{\psi^z_0}$ is gauge-invariant and satisfies $\hat{G}_j\ket{\psi^z_0}=0,\,\forall j$, while $\ket{\psi^x_0}$ is a superposition of superselection sectors and is therefore gauge-noninvariant \cite{SM}. In both states, the left half of the chain is occupied, while the right half is empty, leading to a domain-wall state from the perspective of the matter fields.}
	\label{fig:InitialStates} 
\end{figure}

In a QSM realization of Eq.~\eqref{eq:H0}, unavoidable gauge-breaking errors will arise due to various implementational imperfections \cite{Halimeh2020a}. For $\mathrm{U}(1)$ gauge theories, these can take the form of
\begin{align}\label{eq:H1}
\lambda\hat{H}_1=\lambda\sum_{j=1}^L\bigg[\hat{\sigma}^-_j\hat{\sigma}^-_{j+1}+\hat{\sigma}^+_j\hat{\sigma}^+_{j+1}+\frac{\hat{s}^x_{j,j+1}+\hat{s}^z_{j,j+1}}{\sqrt{S(S+1)}}\bigg],
\end{align}
which, in the language of QED, describes the creation or annihilation of matter without the concomitant change in the electric field required to preserve gauge invariance, or vice versa \cite{Mil2020}. These errors lead to transitions between the different superselection sectors. In addition, Eq.~\eqref{eq:H1} also includes undesired gauge-invariant processes due to $\hat{s}^z_{j,j+1}$. As the main result of our work, we will show that one can efficiently and controllably protect against these errors by employing the experimentally feasible translation-invariant single-body gauge protection \cite{Halimeh2020e}
\begin{align}\nonumber
V\hat{H}_G&=V\sum_j(-1)^j\hat{G}_j\\\label{eq:HG}
&=V\sum_j\big(\hat{n}_j+\hat{s}^z_{j-1,j}+\hat{s}^z_{j,j+1}\big),
\end{align}
which, at large enough $V$, will force the dynamics under the full system $\hat{H}=\hat{H}_0+\lambda\hat{H}_1+V\hat{H}_G$ to be restricted within a set of orthogonal subspaces of the Hilbert space \cite{facchi2002quantum}. In our case, these \textit{quantum Zeno subspaces} are directly connected to the superselection sectors $\mathbf{g}$; see derivational details in Supplemental Material (SM) \cite{SM}. As we show in this Letter, this suppresses transitions between the sectors up to timescales polynomial in $V$, during which DFL is reliably stabilized.
 
For the sake of simplicity, and without loss of generality, we will focus in the main text on the case of $S=1/2$, and relegate results for larger $S$ to the SM \cite{SM}. 

\begin{figure}[t!]
	\centering
	\includegraphics[width=.48\textwidth]{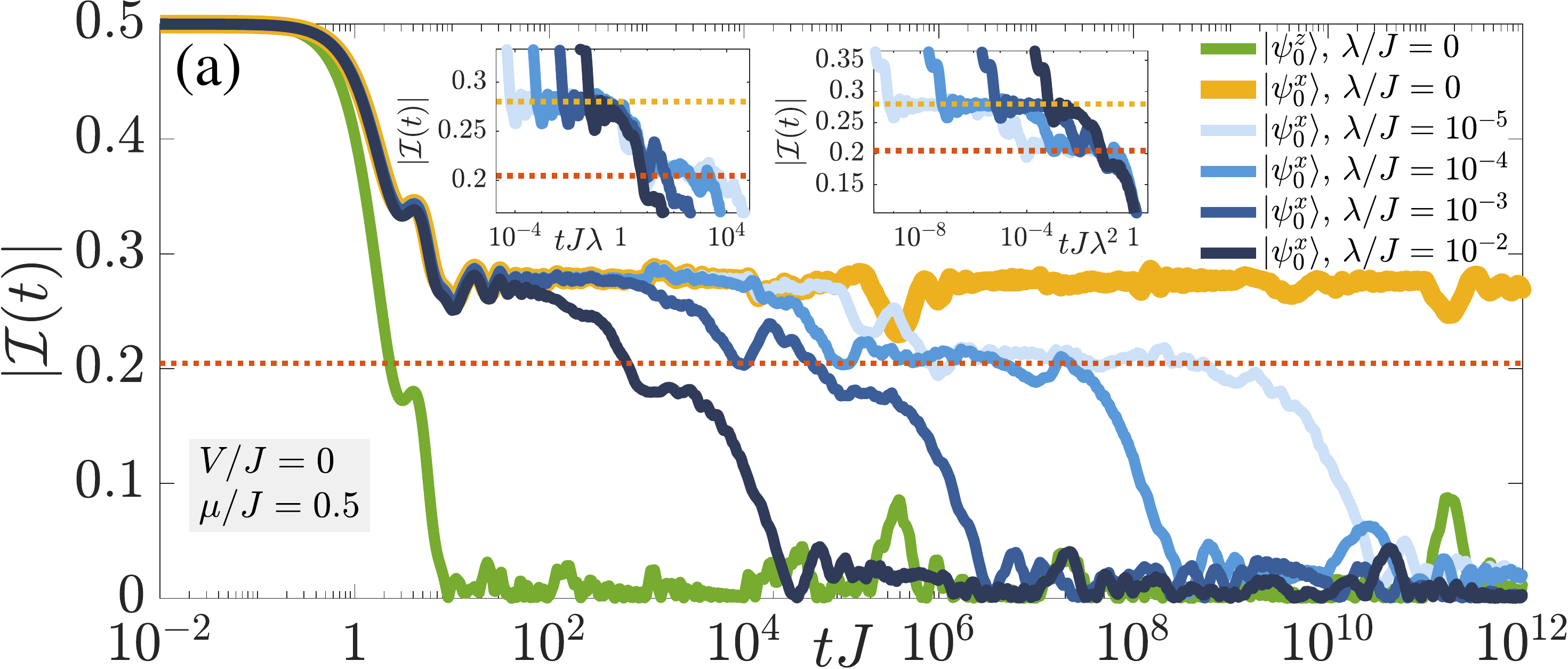}\\
	\vspace{1.1mm}
	\includegraphics[width=.48\textwidth]{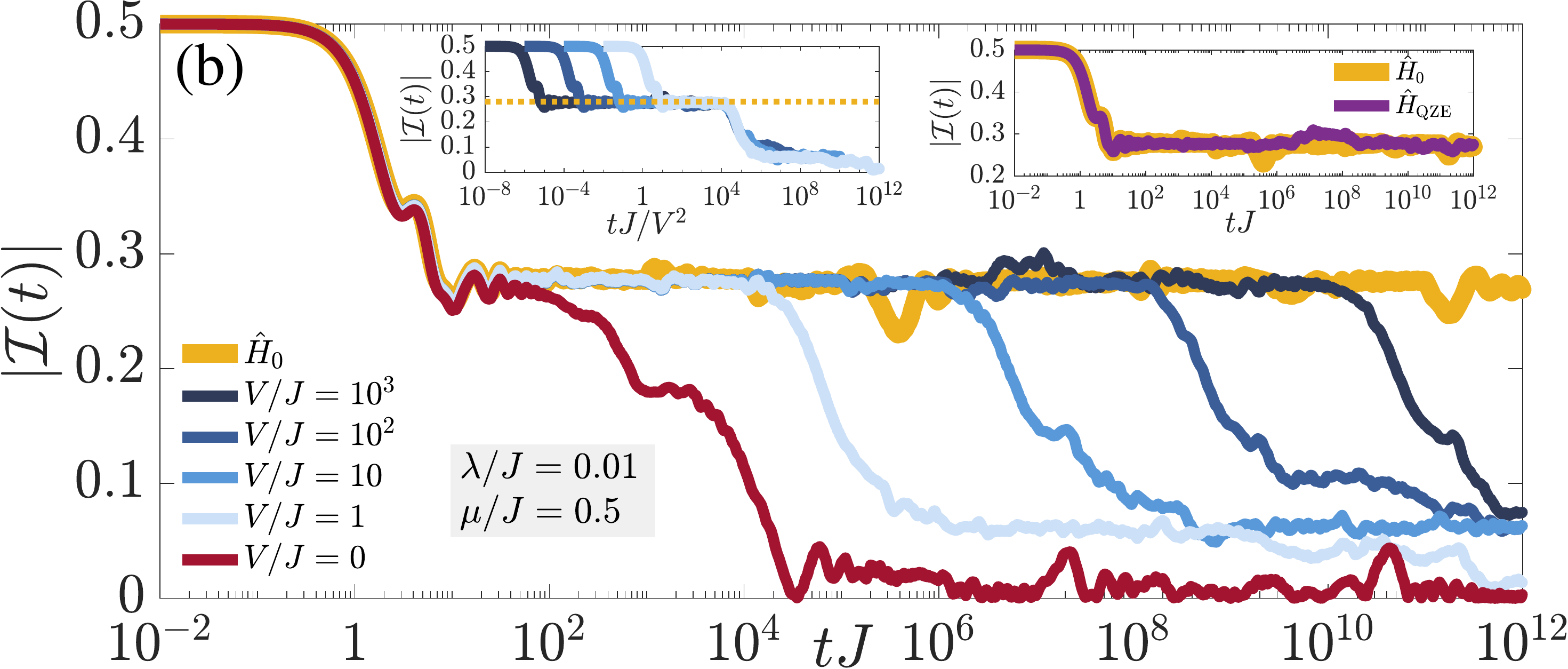}\\
	\vspace{1.1mm}
	\includegraphics[width=.48\textwidth]{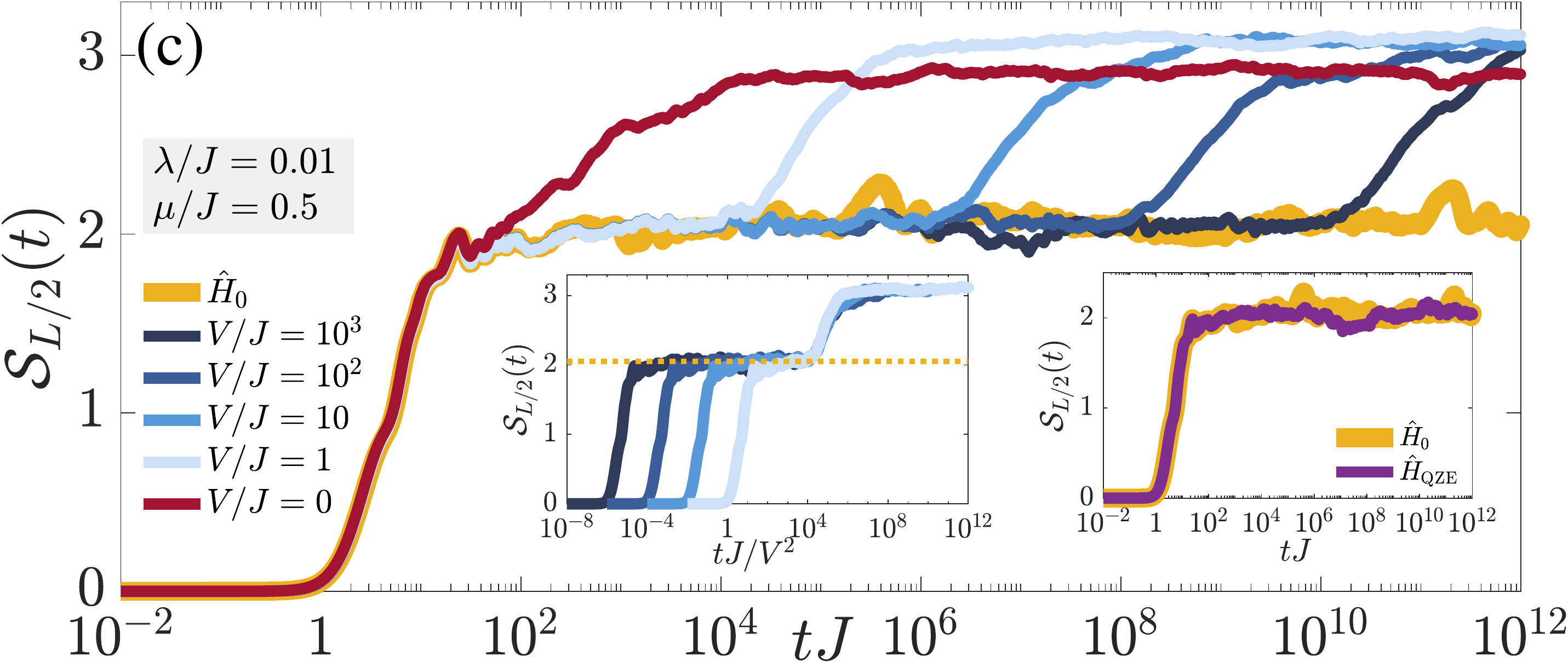}
	\caption{(Color online). Quench dynamics of the imbalance~\eqref{eq:imbalance} calculated through exact diagonalization. (a) Starting in the initial state $\ket{\psi^x_0}$ and quenching with the spin-$1/2$ $\mathrm{U}(1)$ QLM~\eqref{eq:H0} leads to disorder-free localization (yellow) since $\ket{\psi^x_0}$ is a superposition of superselection sectors. In contrast, when the initial state is the gauge-invariant $\ket{\psi^z_0}$ of the superselection sector $g_j=0,\,\forall j$, the quench dynamics under $\hat{H}_0$ shows no localization, and relatively quickly thermalizes to the thermal-ensemble prediction of zero. This also seems to become the case for $\ket{\psi^x_0}$ in the presence of gauge-breaking errors $\lambda\hat{H}_1$ that couple different superselection sectors, where disorder-free localization becomes a prethermal phase lasting up to the timescale $\propto1/\lambda$ (left inset), after which it enters another prethermal plateau up to a timescale $\propto J/\lambda^2$ (right inset), and then thermalizes to zero thereafter. (b) By adding the single-body gauge protection $V\hat{H}_G=\sum_j(-1)^j\hat{G}_j$, we see a restoration of disorder-free localization when starting in $\ket{\psi^x_0}$ up to a timescale $\propto V^2/(\lambda^2J)$ (see left inset), exceeding our analytic predictions. This protection is based on the quantum Zeno effect, where in the $V\to\infty$ limit, the dynamics is exactly reproduced by an emergent gauge theory $\hat{H}_\mathrm{QZE}=\hat{H}_0+\lambda\sum_\mathbf{g}\hat{\mathcal{P}}_\mathbf{g}\hat{H}_1\hat{\mathcal{P}}_\mathbf{g}$ that hosts the same gauge symmetry as $\hat{H}_0$ (see right inset). (c) This behavior is also reflected in the half-chain entanglement entropy, where the single-body gauge protection keeps $\mathcal{S}_{L/2}(t)$ at the plateau of the error-free dynamics up to a timescale $\propto V^2/(\lambda^2J)$. These results are for $L=4$ matter sites with periodic boundary conditions, but our conclusions hold for larger link spin lengths \cite{SM} and system sizes \cite{Z2DFL}.}
	\label{fig:imbalance} 
\end{figure}

\textit{Quench dynamics.--} We consider initial states that are matter domain walls, with the electric fields aligned either along the positive $x$-direction, which we call $\ket{\psi^x_0}$, or along the $z$-direction, denoted as $\ket{\psi^z_0}$. While the latter satisfies $\hat{G}_j\ket{\psi^z_0}=0,\,\forall j$, see Fig.~\ref{fig:InitialStates}, $\ket{\psi^x_0}$ is not gauge-invariant, but rather is in a superposition of superselection sectors. Being product states, both these initial states are straightforward to implement in modern QSM setups.

To probe localization, we calculate in exact diagonalization (ED) the temporally averaged imbalance
\begin{align}\label{eq:imbalance}
\mathcal{I}(t)&=\frac{1}{Lt}\int_0^t ds\sum_{j=1}^Lp_j\bra{\psi(s)}\hat{n}_j\ket{\psi(s)},
\end{align}
where $p_j=\bra{\psi_0}\hat{\sigma}^z_j\ket{\psi_0}$, $\ket{\psi(t)}=e^{-i\hat{H}t}\ket{\psi_0}$, with $\ket{\psi_0}=\ket{\psi^{x,z}_0}$ one of the initial states we consider, and $\hat{H}=\hat{H}_0+\lambda\hat{H}_1+V\hat{H}_G$ is the \textit{faulty} gauge theory. In the case of thermalization, the long-time limit of Eq.~\eqref{eq:imbalance} in the superselection sector $\mathbf{g}=(0,0,0,0)$ is zero. This is indeed the case when the initial state is $\ket{\psi^z_0}$, as can be seen in Fig.~\ref{fig:imbalance}(a) for $\lambda=V=0$ (green curve). In contrast, the quench dynamics of $\ket{\psi^x_0}$ under $\hat{H}_0$ displays localized behavior in the imbalance (yellow), with the latter settling into a nonzero plateau for all accessible times in ED. This DFL has recently been shown to occur in gauge-theory dynamics when the initial state is a superposition of superselection sectors \cite{Smith2017,Brenes2018,Smith2018,Papaefstathiou2020}. Nevertheless, upon turning on $\lambda>0$, but still with $V=0$, this localized phase changes at a timescale $\propto 1/\lambda$ with the imbalance settling into a lower-valued plateau that lasts until a timescale $\propto J/\lambda^2$, after which the imbalance goes to zero, indicating thermalization and absence of localized behavior (shades of blue). This is reminiscent of staircase prethermalization observed in quench dynamics starting in a given superselection sector with small gauge-breaking errors \cite{Halimeh2020b,Halimeh2020c}.

We now employ the single-body gauge protection~\eqref{eq:HG} to stabilize DFL. Starting in the superposition initial state $\ket{\psi^x_0}$ and quenching with the faulty theory $\hat{H}$ with a fixed value of $\lambda=0.01J$, we show in Fig.~\ref{fig:imbalance}(b) the controlled restoration of the DFL phase with increasing $V$. Indeed, even when $V$ is as small as $J$, the DFL plateau is prolonged up to a timescale of over $10^4/J$, which is well beyond the lifetimes accessible in state-of-the-art QSM setups for relevant values of $J\in[10,100]$ Hz \cite{Yang2020}. As $V$ is further increased, the timescale of the DFL plateau is numerically found to be $\propto V^2/(\lambda^2 J)$, which exceeds our analytic predictions \cite{SM}. As we derive analytically based on the QZE \cite{facchi2002quantum,facchi2004unification,facchi2009quantum,burgarth2019generalized,SM}, the quench dynamics under $\hat{H}$ is faithfully reproduced in the large-$V$ limit by the effective Hamiltonian
\begin{align}\label{eq:QZE}
\hat{H}_\mathrm{QZE}=\hat{H}_0+\lambda\sum_\mathbf{g}\hat{\mathcal{P}}_\mathbf{g}\hat{H}_1\hat{\mathcal{P}}_\mathbf{g},
\end{align}
where $\hat{\mathcal{P}}_\mathbf{g}$ is the projector onto the superselection sector $\mathbf{g}$, up to a residual error with upper bound $\propto t V_0^2 /V$, with $V_0$ an energy term that is a linear sum in $\{J,\lambda,\mu,\kappa^2S^2\}$ \cite{SM}. Note that $\hat{H}_\mathrm{QZE}$ is an exact gauge theory with the same gauge symmetry as $\hat{H}_0$, i.e., $[\hat{H}_\mathrm{QZE},\hat{G}_j]=0,\,\forall j$, albeit renormalized due to gauge-invariant processes in $\lambda \hat{H}_1$. The inset of Fig.~\ref{fig:imbalance}(b) shows how $\hat{H}_\mathrm{QZE}$ restores the DFL plateau up to all accessible times in ED. It is worth noting that by removing processes due to $\hat{s}^z_{j,j+1}$ in $\lambda\hat{H}_1$, then the term $\hat{\mathcal{P}}_\mathbf{g}\hat{H}_1\hat{\mathcal{P}}_\mathbf{g}=0$, which renders $\hat{H}_\mathrm{QZE}=\hat{H}_0$, and the DFL plateau is restored quantitatively in addition to qualitatively \cite{SM}.

We emphasize that the protection term~\eqref{eq:HG} is translation-invariant and has neither quenched disorder nor spatial inhomogeneity. As such, the restoration of the DFL phase by this term cannot be attributed to disorder-MBL \cite{Basko2006} or Stark-MBL \cite{Schulz2019}. Indeed, employing this term in the quench dynamics of $\ket{\psi^z_0}$ leads to no localized dynamics even when $\lambda=0$ \cite{SM}. Moreover, we have also checked that a thermal-ensemble prediction will always give a zero imbalance for a quench starting in $\ket{\psi^x_0}$, which rules out that the restored plateau due to gauge protection is thermal \cite{SM}. To firm up this picture, we calculate the half-chain entanglement entropy $\mathcal{S}_{L/2}(t)$, shown in Fig.~\ref{fig:imbalance}(c). The dynamics of this quantity under gauge protection reproduces the ideal error-free entanglement entropy up to a timescale $\propto V^2/(\lambda^2 J)$, indicating restoration of truly localized dynamics.

\begin{figure}[t!]
	\centering
	\includegraphics[width=.48\textwidth]{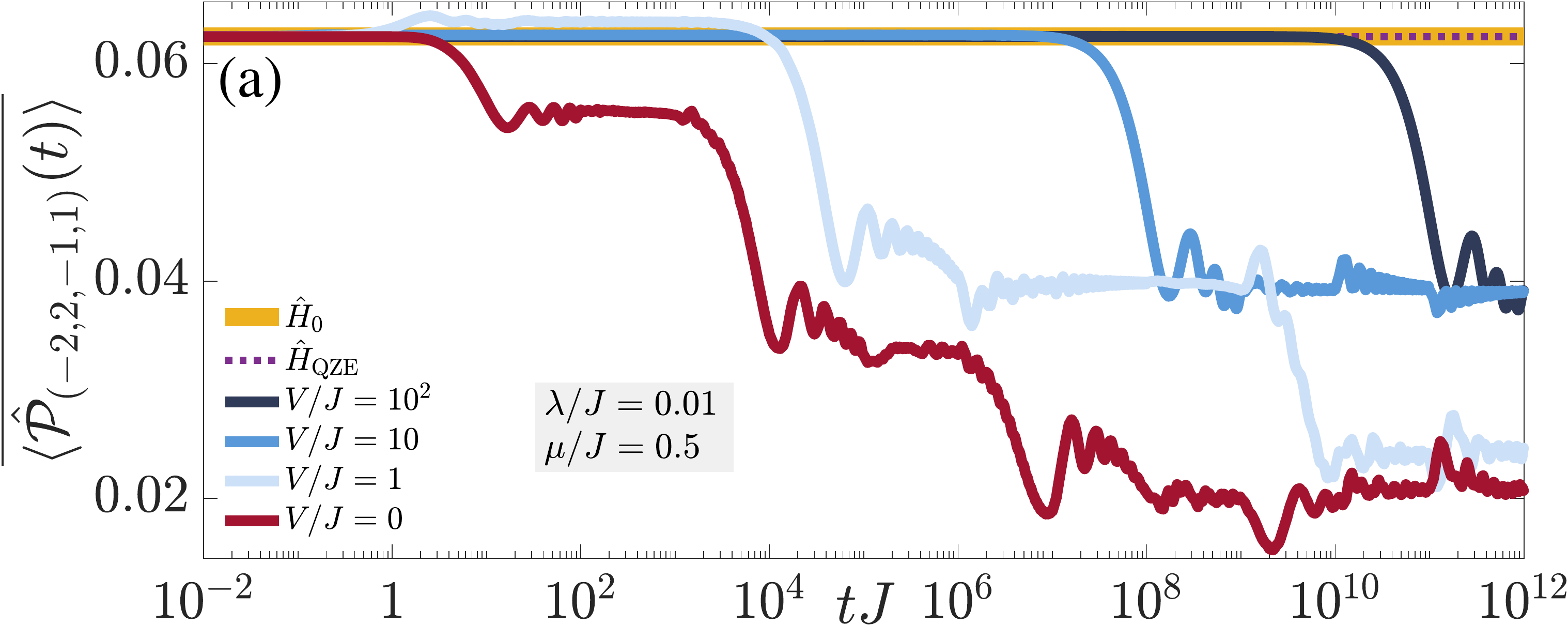}\\
	\vspace{1.1mm}
	\includegraphics[width=.48\textwidth]{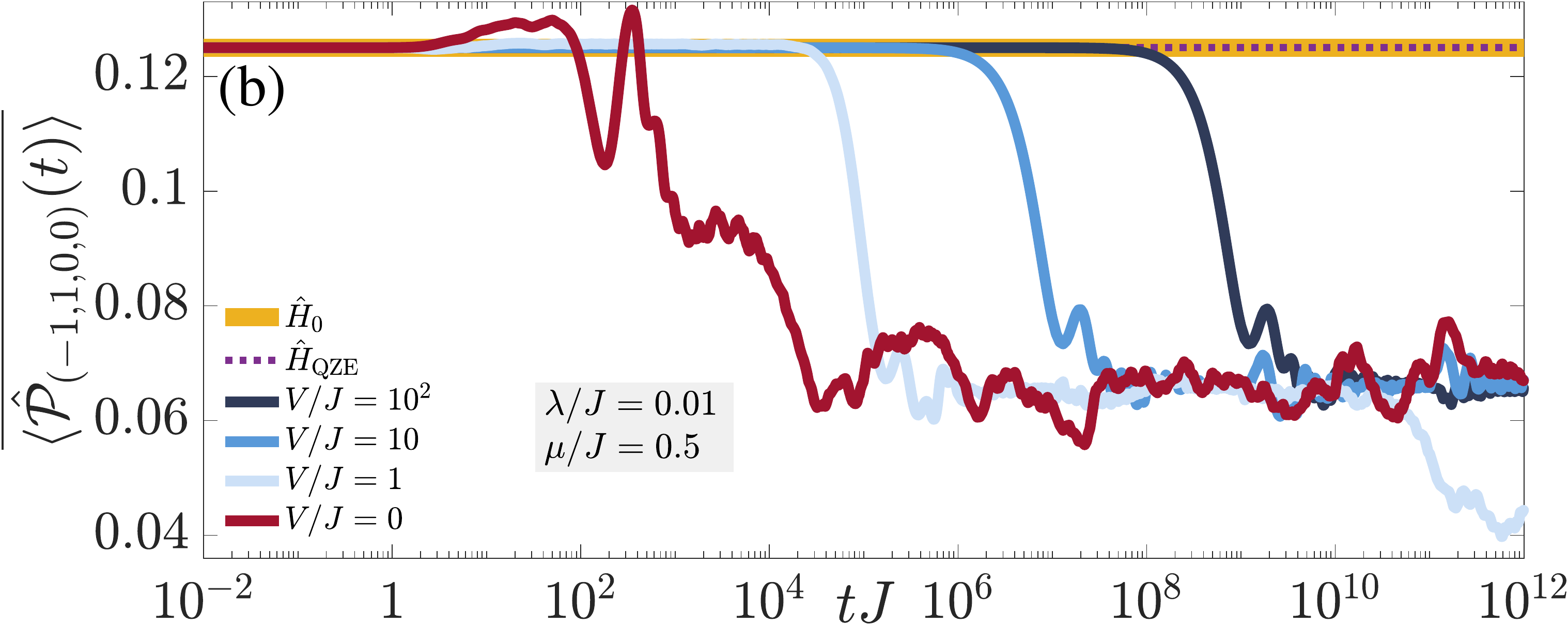}
	\caption{(Color online). Quench dynamics of the projectors $\hat{\mathcal{P}}_\mathbf{g}$ onto the superselection sectors (a) $\mathbf{g}=(-2,2,-1,1)$ and (b) $\mathbf{g}=(-1,1,0,0)$ starting in $\ket{\psi^x_0}$. In the ideal case ($\lambda=V=0$), the expectation value of $\hat{\mathcal{P}}_\mathbf{g}$ is unchanged throughout the whole time evolution (yellow curve), and so the corresponding quench dynamics of the imbalance~\eqref{eq:imbalance} is purely due to intra-sector processes (see yellow curve in Fig.~\ref{fig:imbalance}). When gauge-breaking errors are turned on ($\lambda>0$), $\langle\hat{\mathcal{P}}_\mathbf{g}\rangle$ begins to drastically deviate at early times from its initial value, indicating transitions between different superselection sectors. Upon adding single-body gauge protection ($V>0$), $\langle\hat{\mathcal{P}}_\mathbf{g}\rangle$ persists longer at its initial value, up to a timescale that is at least $\propto V^2/(\lambda^2J)$, during which inter-sector dynamics is suppressed, thereby restoring disorder-free localization.}
	\label{fig:projectors}
\end{figure}

To further understand how single-body gauge protection stabilizes DFL, let us look at the temporally averaged expectation values of the projectors $\hat{\mathcal{P}}_\mathbf{g}$. The initial state $\ket{\psi^x_0}$ is a superposition over $15$ out of $134$ possible superselection sectors for the model we use. We show in Fig.~\ref{fig:projectors}(a,b) the quench dynamics for the sectors $(-2,2,-1,1)$ and $(-1,1,0,0)$, respectively, although our conclusions are the same for other sectors. In the absence of gauge-breaking errors, $\langle\hat{\mathcal{P}}_\mathbf{g}(t)\rangle$ for any $\mathbf{g}$ remains constant at all accessible times in ED, indicating no inter-sector dynamics. In contrast, in the presence of errors and without protection, $\langle\hat{\mathcal{P}}_\mathbf{g}(t)\rangle$ deviates quickly and significantly from its initial value (red curves) due to transitions between different superselection sectors facilitated by $\lambda\hat{H}_1$. In this case, the time-evolved projected wave function spreads over all $134$ possible sectors, and its dynamics is nevertheless no longer localized. This may seem counterintuitive, but makes sense because DFL arises due to \textit{both} starting in a superposition of superselection sectors \textit{and} the absence of inter-sector coupling. In the presence of inter-sector coupling, the background charges are no longer fixed, which can be interpreted as a time-varying disorder potential, which is known to lift localization \cite{Medvedyeva2016,Fischer2016,Levi2016,Smith2018,Maier2019}.

Upon adding the single-body gauge protection~\eqref{eq:HG}, $\langle\hat{\mathcal{P}}_\mathbf{g}(t)\rangle$ stays longer at its initial value up to a timescale $\propto V^2/(\lambda^2 J)$, which translates to a suppression of inter-sector dynamics within this timescale (Fig.~\ref{fig:projectors}, shades of blue). The behavior of the superselection-sector projectors gives us further insight into the workings of single-body gauge protection and how it restores DFL. The addition of $V\hat{H}_G$ at sufficiently large $V$ constrains the dynamics to each superselection sector while suppressing transitions between these sectors in congruence with the QZE. This is a necessary ingredient for DFL \cite{Smith2017,Brenes2018}, which gauge-breaking errors inevitably destroy, but single-body gauge protection ultimately restores up to impressive timescales that match or exceed typical experimental lifetimes of modern QSM setups.

\textit{Summary.--} In this Letter, we have presented single-body gauge protection as a viable and powerful tool for the stabilization of DFL in gauge-theory dynamics subjected to experimentally relevant gauge-breaking errors. The protection term is fully translation-invariant and spatially homogeneous, and does not on its own lead to localized behavior. It induces quantum Zeno dynamics suppressing transitions between different superselection sectors and restricting the time evolution to intra-sector dynamics. 

We have presented numerical evidence obtained from ED to demonstrate the restoration of DFL in the imbalance and mid-chain entanglement entropy when quenching from a simple initial state which corresponds to a superposition over an extensive number of gauge sectors. Additionally, we have shown through the dynamics of projectors onto superselection sectors how single-body gauge protection suppresses inter-sector processes, extending the timescale over which the expectation value of the projector remains at its initial value, and therefore protecting the localized dynamics.

Even though our analytic predictions yield a worst-case timescale linear in the protection strength, our ED calculations suggest that in fact this timescale is at least quadratic in the protection strength. For even moderate values of the latter equal to the coupling constant $J$ we find localization timescales of roughly $10^4/J$, which competes with the lifetimes achieved in the most advanced quantum simulator platforms \cite{Bernien2017,Yang2020,Zhou2021}. From an experimental perspective, our novel protection scheme has the advantage that it only needs single-body single-species terms in the $z$-basis, which are much simpler to implement than the ideal gauge theory itself. Given the difficulty of implementing error-free gauge theories with dynamical matter and gauge fields in current quantum synthetic matter setups \cite{Schweizer2019,Mil2020,Yang2020}, the impressive efficacy of single-body gauge protection and its experimental feasibility will considerably facilitate the observation of DFL.

Finally, we note that even though in the main text we have focused on the spin-$1/2$ $\mathrm{U}(1)$ quantum link model, our method applies to any spin length $S$ (see SM \cite{SM}), and can be readily extended to any generic Abelian gauge theory \cite{Z2DFL}. Moreover, due to the weak dependence of the quantum Zeno effect on system size, we expect our results to be valid for large-scale systems \cite{SM,Z2DFL}.

\begingroup
\renewcommand{\addcontentsline}[3]{}
\renewcommand{\section}[2]{}
\begin{acknowledgments}
J.C.H.~is grateful to Haifeng Lang for fruitful discussions and valuable comments. This work is part of and supported by Provincia Autonoma di Trento, the ERC Starting Grant StrEnQTh (project ID 804305), the Google Research Scholar Award ProGauge, and Q@TN — Quantum Science and Technology in Trento.  H.Z.~acknowledges support from a Doctoral-Program Fellowship of
the German Academic Exchange Service (DAAD). We acknowledge support from the Imperial--TUM flagship partnership.
\end{acknowledgments}

\endgroup

\clearpage
\pagebreak
\newpage
\setcounter{equation}{0}
\setcounter{figure}{0}
\setcounter{table}{0}
\setcounter{page}{1}
\makeatletter
\renewcommand{\bibnumfmt}[1]{[S#1]}
\renewcommand{\citenumfont}[1]{S#1}
\renewcommand{\theequation}{S\arabic{equation}}
\renewcommand{\thefigure}{S\arabic{figure}}
\renewcommand{\thetable}{S\Roman{table}}
\renewcommand{\bibnumfmt}[1]{[S#1]}
\renewcommand{\citenumfont}[1]{S#1}
\begin{center}
\textbf{--- Supplemental Material ---\\Stabilizing Disorder-Free Localization}
\end{center}
\tableofcontents

\section{Supporting numerical results for the spin-$1/2$ $\mathrm{U}(1)$ quantum link model}
We provide here supporting numerical results to further confirm our conclusions from the main text, illustrate the efficacy of single-body gauge protection, and highlight the independence of our results of the specific parameter choices used in the main text. As in the main text, we will consider the initial states shown in Fig.~\ref{fig:InitialStates}. The gauge-invariant superselection sectors spanned by these initial states are shown in Table~\ref{TableG}.

\begin{figure*}[t!]
	\centering
	\includegraphics[width=.48\textwidth]{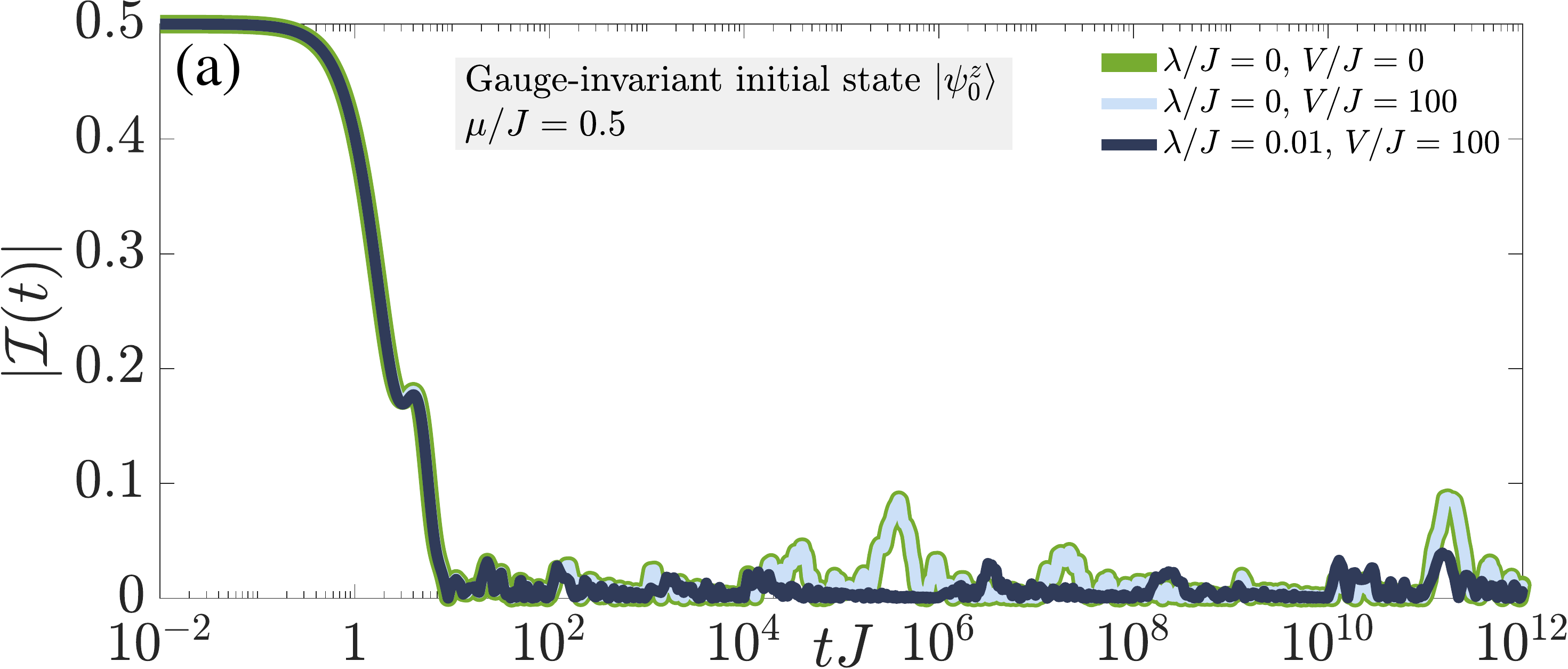}\quad\includegraphics[width=.48\textwidth]{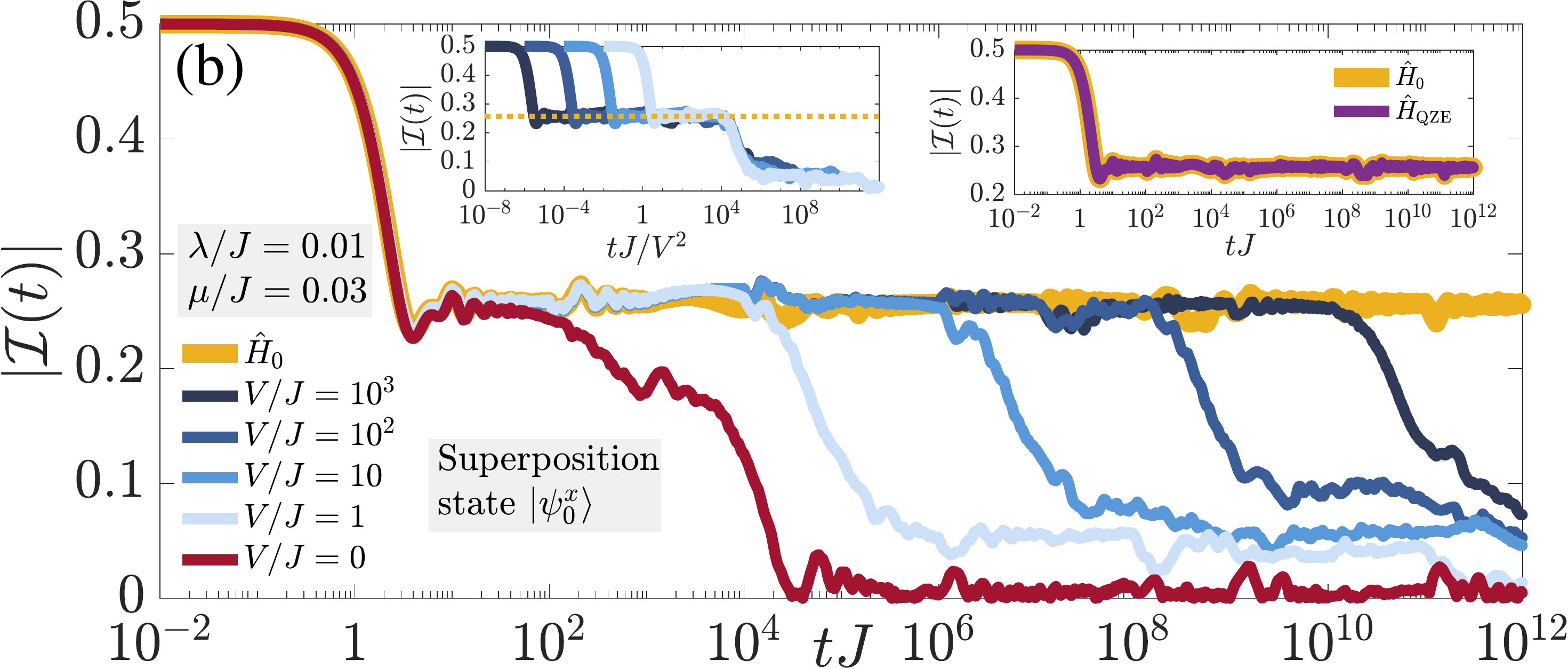}\\
	\vspace{1.1mm}
	\includegraphics[width=.48\textwidth]{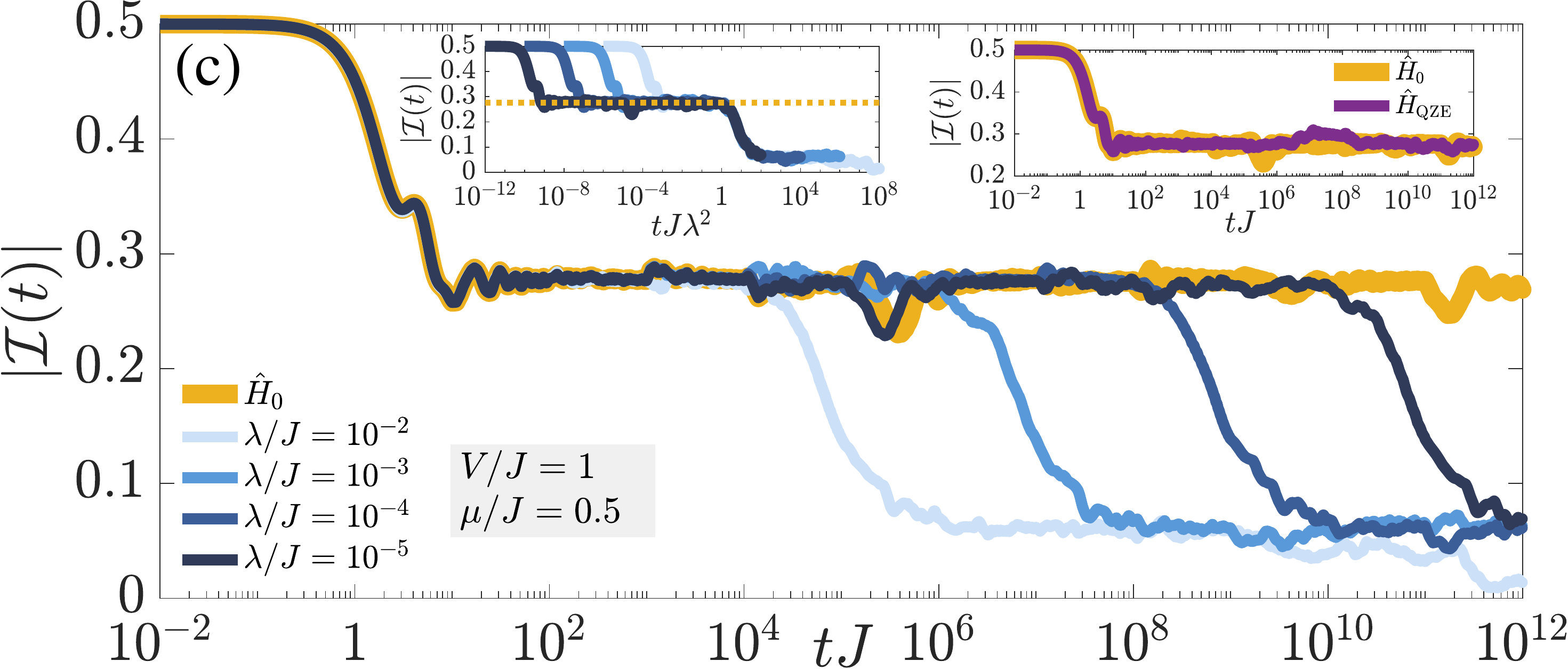}\quad\includegraphics[width=.48\textwidth]{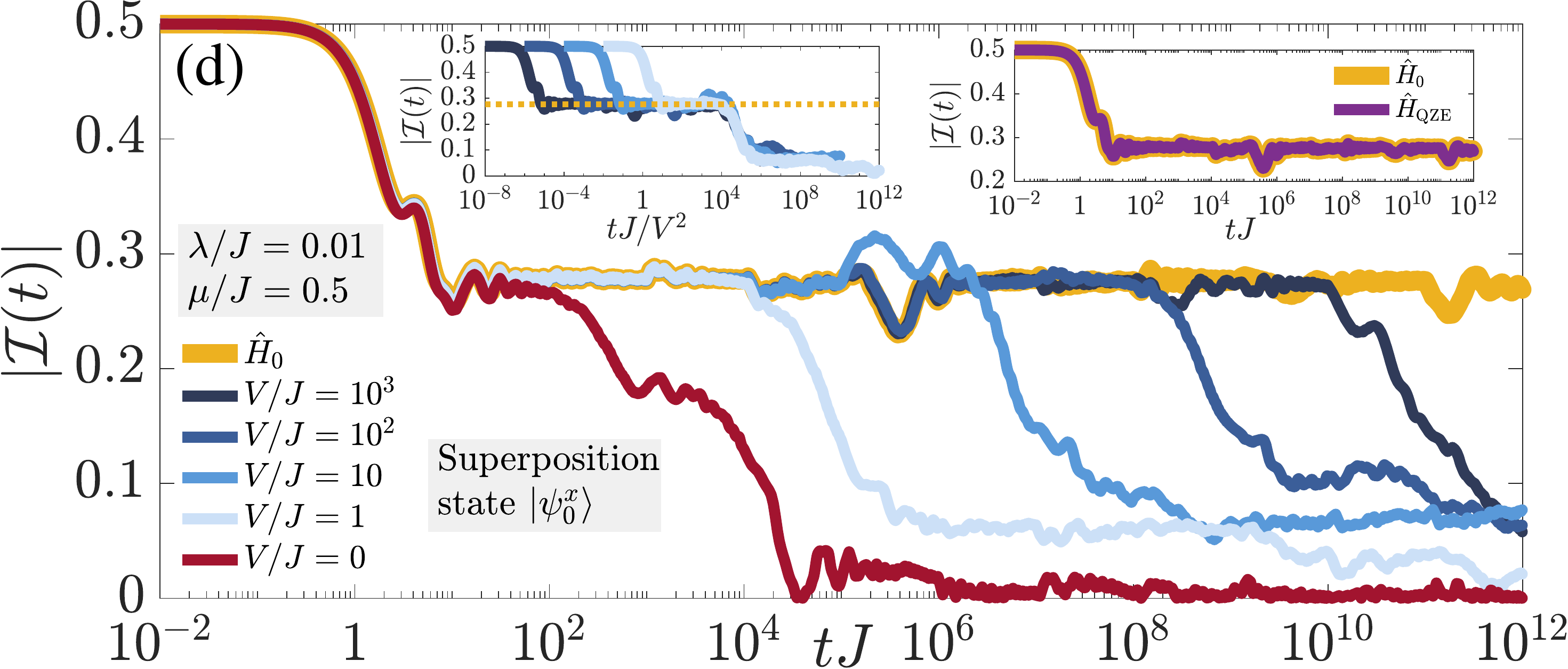}
	\caption{(Color online). Supporting numerical results for the spin-$1/2$ $\mathrm{U}(1)$ quantum link model. (a) Starting in the gauge-invariant initial state $\ket{\psi^z_0}$ (see Fig.~\ref{fig:InitialStates}) and quenching with $\hat{H}=\hat{H}_0+\lambda\hat{H}_1+V\hat{H}_G$ will not lead to any localized dynamics in the imbalance. This shows that single-body gauge protection does not induce disorder-free localization on its own, but rather this requires a superposition initial state as in the ideal case. Gauge protection merely protects disorder-free localization. (b) Same as Fig.~\ref{fig:imbalance}(b) but with $\mu=0.03J$. The qualitative conclusions remain the same, showing that the value of $\mu$ is not relevant to our findings. (c) Same as Fig.~\ref{fig:imbalance}(b) but here we fix $V=J$ while varying $\lambda$, verifying the timescale of the emergent gauge theory to be $\propto V^2/(\lambda^2J)$. (d) Same as Fig.~\ref{fig:imbalance}(b) but where the error term is now $\lambda\hat{H}_1'$ given in Eq.~\eqref{eq:H1p}. Since $\hat{P}_\mathbf{g}\hat{H}_1'\hat{P}_\mathbf{g}=0,\,\forall\mathbf{g}$, in the large-$V$ limit the emergent gauge theory will be exactly equal to the ideal theory: $\hat{H}_\mathrm{QZE}=\hat{H}_0$ (see right inset). This leads to the quantitative, in addition to qualitative, reproduction of the ideal dynamics when $V$ is large enough.}
	\label{fig:supp}
\end{figure*}

To start with, we look at the effect of single-body gauge protection on the quench dynamics of the gauge-invariant initial state $\ket{\psi^z_0}$ (see Fig.~\ref{fig:InitialStates}). As shown in Fig.~\ref{fig:supp}(a), the imbalance will always go to zero upon quenching $\ket{\psi^z_0}$ with the faulty theory $\hat{H}=\hat{H}_0+\lambda\hat{H}_1+V\hat{H}_G$ regardless of the value of $\lambda$ or $V$. This demonstrates that the emergence of localized behavior in quench dynamics propagated by $\hat{H}$ still depends on the initial state. In other words, gauge protection only preserves disorder-free localization if it is already existent in the error-free quench protocol. 

Until now, we have focused on $\mu=0.5J$ for the mass value, but this choice is not special, and our conclusions hold for other values of the mass, as shown in Fig.~\ref{fig:supp}(b) for $\mu=0.03J$.

In order to verify that the timescale of the emergent gauge theory is inversely proportional to the error strength, we fix $V=J$ and scan $\lambda$ over several values in Fig.~\ref{fig:supp}(c). The timescale is clearly proportional to $V^2/(\lambda^2 J)$ (see inset). This scaling lies in the quantum Zeno dynamics, which by definition excludes first-order transitions due to $\lambda H_1$, see Eq.~\eqref{eq:QZE}. As such, when the regime of quantum Zeno dynamics fails, it is when second-order processes in $\lambda H_1$ dominate, giving a timescale inversely proportional to the error strength when single-body gauge protection is turned on.

\begin{table}[t!]
	\centering
	\begin{tabular}{|| c || c | c ||}
		\hline
		 $\mathbf{g}=(g_1,g_2,g_3,g_4)$ & $\bra{\psi^z_0}\hat{\mathcal{P}}_\mathbf{g}\ket{\psi^z_0}$ & $\bra{\psi^x_0}\hat{\mathcal{P}}_\mathbf{g}\ket{\psi^x_0}$ \\ [0.5ex] 
		\hline\hline
		$(-2,1,0,0)$ & $0$ & $0.0625$\\ 
		\hline
		$(-2,1,1,0)$ & $0$ & $0.0625$\\
		\hline
		$(-2,2,-1,1)$ & $0$ & $0.0625$\\
		\hline
		$(-2,2,0,0)$ & $0$ & $0.0625$\\
		\hline
		$(-1,0,0,1)$ & $0$ & $0.0625$\\
		\hline
		$(-1,0,1,0)$ & $0$ & $0.0625$\\
		\hline
		$(-1,1,-1,1)$ & $0$ & $0.0625$\\
		\hline
		$(-1,1,0,0)$ & $0$ & $0.125$\\
		\hline
		$(-1,1,1,-1)$ & $0$ & $0.0625$\\
		\hline
		$(-1,2,-1,0)$ & $0$ & $0.0625$\\
		\hline
		$(-1,2,0,-1)$ & $0$ & $0.0625$\\
		\hline
		$(0,0,0,0)$ & $1$ & $0.0625$\\
		\hline
		$(0,0,1,-1)$ & $0$ & $0.0625$\\
		\hline
		$(0,1,-1,0)$ & $0$ & $0.0625$\\
		\hline
		$(0,1,0,-1)$ & $0$ & $0.0625$\\ [1ex] 
		\hline
	\end{tabular}
	\caption{The gauge-invariant superselection sectors spanned by the gauge-invariant initial state $\ket{\psi^z_0}$ and the superposition initial state $\ket{\psi_0^x}$ shown in Fig.~\ref{fig:InitialStates}, with $\ket{\psi^z_0}$ residing only in $\mathbf{g}=(0,0,0,0)$, while $\ket{\psi^x_0}$ is a superposition of several superselection sectors. Here, $\hat{\mathcal{P}}_\mathbf{g}$ is the projector onto the superselection sector $\mathbf{g}$, and in this Table we consider a spin-$1/2$ $\mathrm{U}(1)$ quantum link model on a lattice of $L=4$ matter sites and $L=4$ gauge links with periodic boundary conditions.}
	\label{TableG}
\end{table}

In the main text, we have considered gauge-breaking errors given by Eq.~\eqref{eq:H1}. The presence of the electric-field operator $\hat{s}^z_{j,j+1}$ leads to gauge-invariant processes due to $\lambda\hat{H}_1$ in addition to gauge-noninvariant ones. This means that $\sum_\mathbf{g}\hat{P}_\mathbf{g}\hat{H}_1\hat{P}_\mathbf{g}\neq0$, and leads to a renormalization of the gauge theory in the large-$V$ limit as $\hat{H}_\mathrm{QZE}=\hat{H}_0+\lambda\sum_\mathbf{g}\hat{P}_\mathbf{g}\hat{H}_1\hat{P}_\mathbf{g}$. Let us instead consider the error term
\begin{align}\label{eq:H1p}
\lambda\hat{H}_1'=\lambda\sum_{j=1}^L\bigg[\hat{\sigma}^-_j\hat{\sigma}^-_{j+1}+\hat{\sigma}^+_j\hat{\sigma}^+_{j+1}+\frac{\hat{s}^x_{j,j+1}}{\sqrt{S(S+1)}}\bigg],
\end{align}
which satisfies $\hat{P}_\mathbf{g}\hat{H}_1\hat{P}_\mathbf{g}=0,\,\forall\mathbf{g}$, i.e., it includes purely gauge-noninvariant processes. As shown in Fig.~\ref{fig:supp}(d), quenching the superposition state $\ket{\psi^x_0}$ with the faulty theory $\hat{H}_0+\lambda\hat{H}_1'+V\hat{H}_G$ leads to the qualitative as well as quantitative restoration of disorder-free localization. This makes sense since in the large-$V$ limit the emergent gauge theory is $\hat{H}_\mathrm{QZE}=\hat{H}_0$, i.e., exactly the ideal gauge theory.

\begin{figure*}[t!]
	\centering
	\includegraphics[width=.48\textwidth]{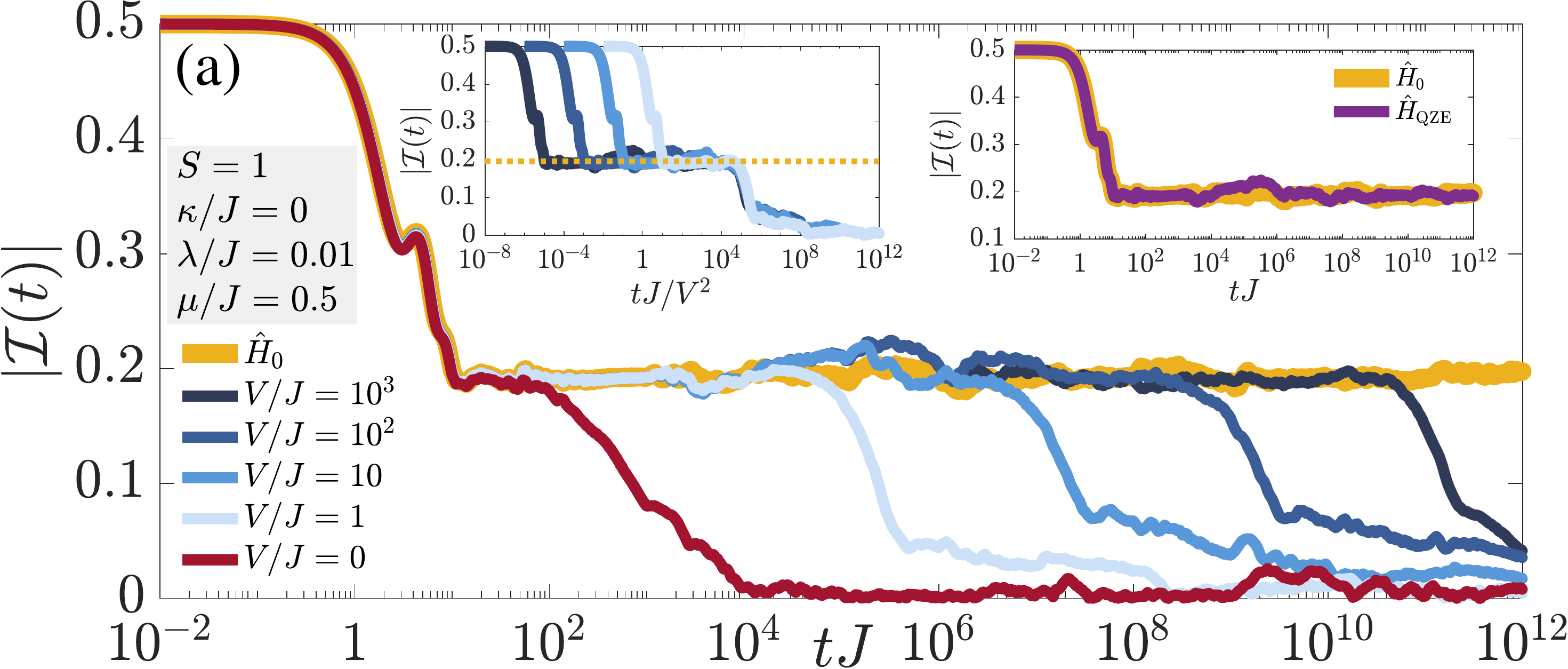}\quad
	\includegraphics[width=.48\textwidth]{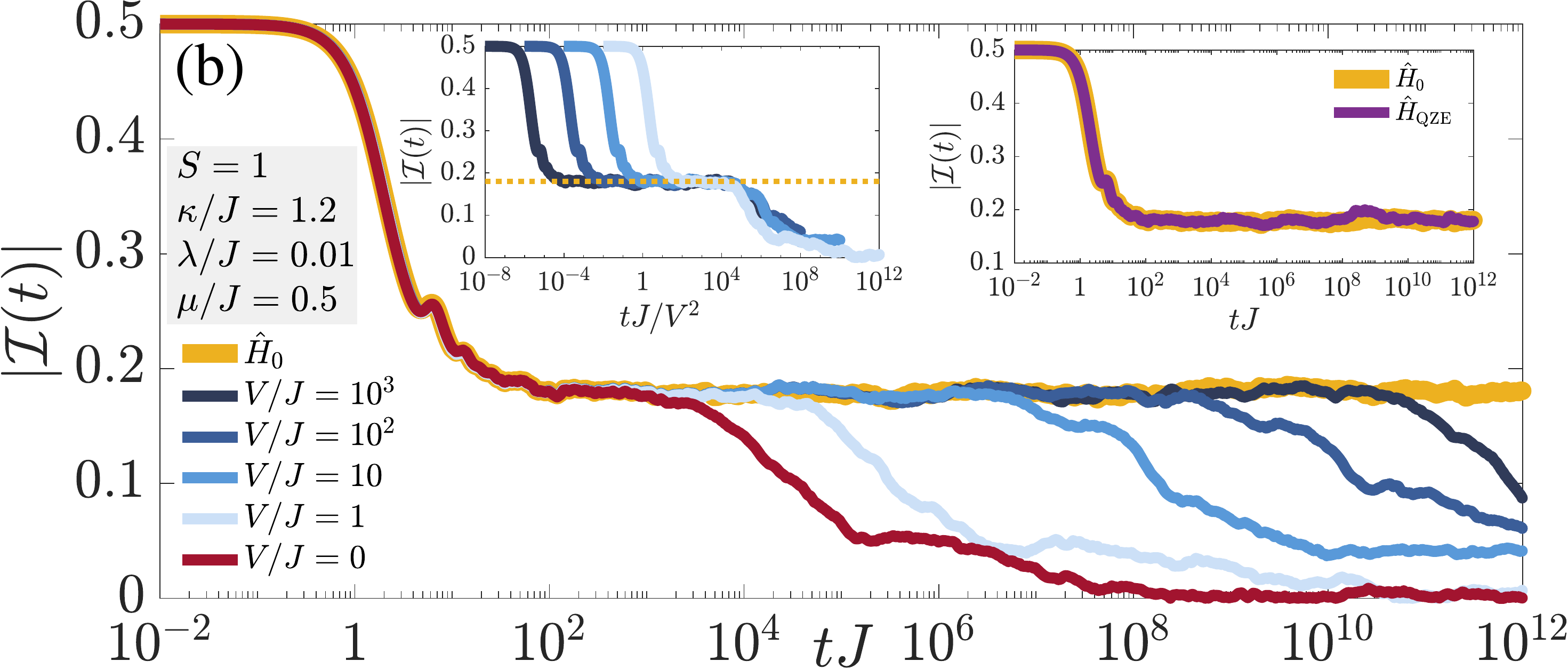}
	\vspace{1.1mm}
	\includegraphics[width=.48\textwidth]{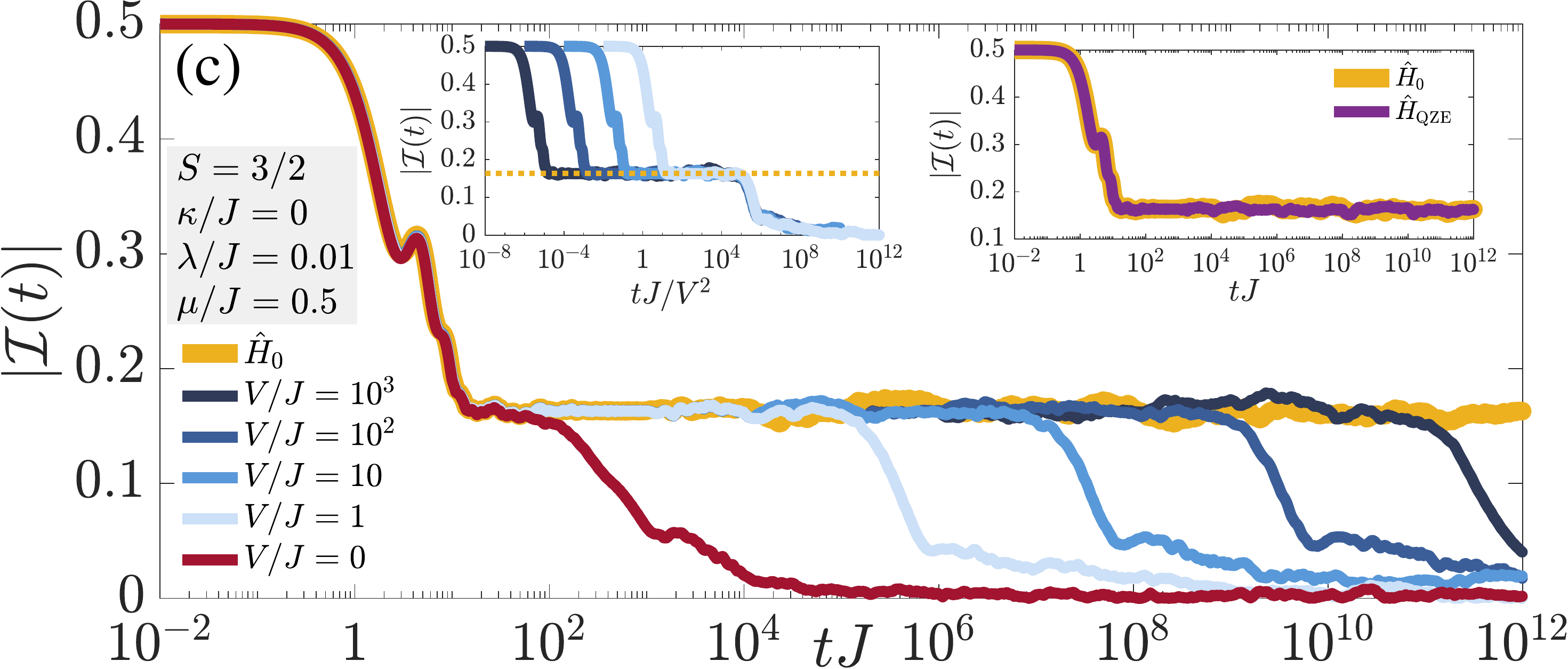}\quad
	\includegraphics[width=.48\textwidth]{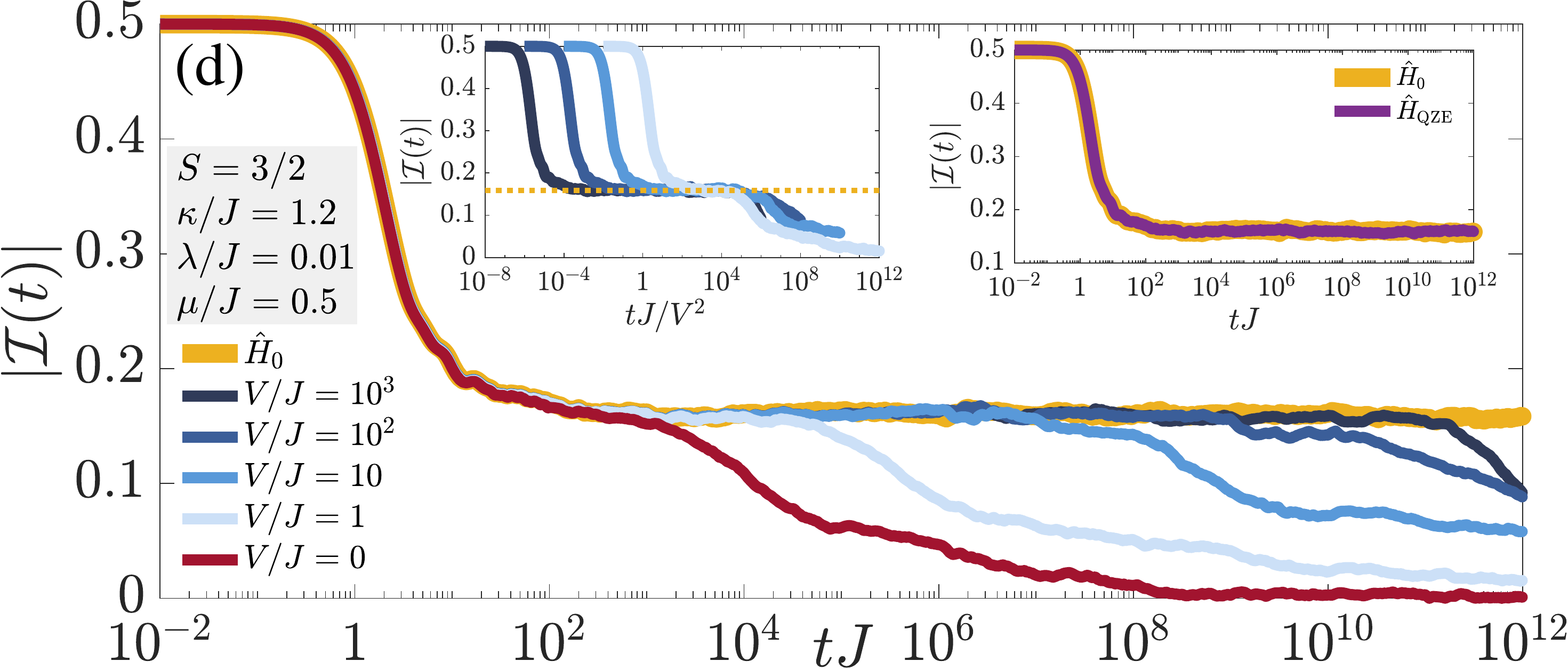}
	\caption{(Color online). Same as Fig.~\ref{fig:imbalance}(b) but for the (a,b) spin-$1$ and (c,d) spin-$3/2$ $\mathrm{U}(1)$ quantum link models for (a,c) $\kappa=0$ and (b,d) $\kappa=1.2J$. As can be seen, the qualitative conclusions are identical to the spin-$1/2$ case.}
	\label{fig:imbalance_largerSpin} 
\end{figure*}

\section{Spin-$S$ $\mathrm{U}(1)$ quantum link model with $S>1/2$}
Our conclusions hold for larger link spin lengths $S$. As an example, we consider the same quench of the superposition state $\ket{\psi^x_0}$ by the faulty theory $\hat{H}=\hat{H}_0+\lambda\hat{H}_1+V\hat{H}_G$, the dynamics of which in the case of $S=1/2$ are shown in Fig.~\ref{fig:imbalance}(b), and repeat it for $S=1$ and $S=3/2$ in Fig.~\ref{fig:imbalance_largerSpin}. We consider two values of the electric-field coupling strength $\kappa=0$ and $1.2$, which is irrelevant in the spin-$1/2$ case since then $(\hat{s}^z_{j,j+1})^2=\hat{\mathds{1}}_{j,j+1}$ leads to an inconsequential energy constant. In all cases, we see that the conclusions are identical to those of the spin-$1/2$ case, with single-body gauge protection restoring disorder-free localization up to a timescale $\propto V^2/(\lambda^2 J)$.

\section{Thermal ensemble}
As discussed in the main text, we have calculated the corresponding canonical thermal ensemble for the quenches considered in this work. This is done by requiring that
\begin{align}\label{eq:energy}
\bra{\psi_0}\hat{H}\ket{\psi_0}\mbeq\Tr\big\{\hat{H}\hat{\rho}_\mathrm{CE}\big\},
\end{align}
where the canonical thermal ensemble is
\begin{align}
\hat{\rho}_\mathrm{CE}=\frac{e^{-\beta\hat{H}}}{\Tr\big\{e^{-\beta\hat{H}}\big\}}.
\end{align}
The enforcement of Eq.~\eqref{eq:energy} is necessary due to energy conservation throughout the unitary dynamics. The only unknown in Eq.~\eqref{eq:energy} is the inverse temperature $\beta$, which can be computed using Newton's method, for example. Once $\beta$, and hence $\hat{\rho}_\mathrm{CE}$, are determined, the thermal-ensemble prediction of any local observable can be attained. For all the quenches considered in this work, we find that the thermal ensemble always predicts a zero imbalance, including when single-body gauge protection is employed. As such, the plateau restored by gauge protection is not thermal, but rather a signature of localized dynamics.

\section{Quantum Zeno effect}
As mentioned in the main text, the quantum Zeno effect plays a principal role in our single-body gauge protection scheme, as we will explain here. Even though the quantum Zeno effect is also used in the context of open systems, in our discussion we will focus on the coherent case.

Quantum Zeno dynamics \cite{facchi2002quantum-S,facchi2004unification-S,facchi2009quantum-S,burgarth2019generalized-S} has different manifestations. The traditional picture of the quantum Zeno effect is based on \textit{pulsed measurements}, which stems from von Neumann's projection postulate \cite{von1955mathematical-S}. One performs frequent projective measurements to restrict the dynamics within the sectors of the measurement. In a \textit{selective} measurement, the projector operator representing the measurement will project onto a \textit{selected} subsector of the Hilbert space containing the initial state, or onto the initial state itself. However, one can also employ the concept of \textit{nonselective} measurements, where the projector is onto orthogonal subspaces of the Hilbert space, rather than a single subspace. Concretely, let us consider a Hamiltonian $\hat{H}$ and assume the system is initialized in an initial state $\hat{\rho}_0=\ketbra{\psi_0}$. Consider an evolution time $t_f$ during which we make $N\to\infty$ measurements on the system at equal time intervals $\delta t=t_f/N$. Let the set of orthogonal projectors $\hat{P}_m$ denote these nonselective measurements. These projectors satisfy the properties $\hat{P}_k\hat{P}_l=\hat{P}_k\delta_{k,l}$ and $\sum_k\hat{P}_k=\hat{\mathds{1}}$. One can then formally show that in the $N\to\infty$ limit \cite{facchi2009quantum-S}
\begin{align}\nonumber
    \hat{\rho}(t_f)&=\lim_{N\to\infty}\sum_k\hat{U}_k(t_f)\hat{\rho}_0\hat{U}_k^\dagger(t_f),
\end{align}
where $\hat{U}_k(t)=\hat{P}_k e^{-i\hat{P}_k\hat{H}\hat{P}_kt}$.
As such, we see that in the limit of infinitely frequent measurements, the dynamics is propagated by the effective Hamiltonian $\hat{H}_\mathrm{QZE}=\sum_k\hat{P}_{k}\hat{H}\hat{P}_{k}$, which restricts the dynamics to within each subsector $k$ of projector $\hat{P}_k$.

However, pulsed measurements are not the only mechanism by which quantum Zeno dynamics is induced. One can equally well introduce a \textit{strong continuous coupling} in the dynamics of the system, which will lead to the creation of quantum Zeno subspaces \cite{facchi2002quantum-S}, mimicking pulsed measurements. Let us now assume our Hamiltonian is $\hat{H}=\hat{H}_0+\lambda\hat{H}_1+V\hat{H}_G$ as in the main text, but, for generality, we take $\hat{H}_G=\sum_jc_j\hat{G}_j$. We will now prove, following the steps detailed in Ref.~\cite{facchi2002quantum-S}, that in the infinite-$V$ limit the time-evolution operator $\hat{U}_V(t)=e^{-i\hat{H}t}$ is diagonal relative to $\hat{H}_G$. Specifically, we will show that
\begin{align}
    \big[\hat{U}_{V\to\infty}(t),\hat{P}_\alpha\big]=0,
\end{align}
where $\hat{H}_G\hat{P}_\alpha=\epsilon_\alpha\hat{P}_\alpha$, and $\hat{P}_\alpha$ is the orthogonal projection onto the eigenspace $\mathcal{H}_{\hat{P}_\alpha}$ of $\hat{H}_G$ with eigenvalue $\epsilon_\alpha$. This spectral decomposition of $\hat{H}_G$ is nondegenerate, i.e., if $\alpha\neq\alpha'\iff\epsilon_\alpha\neq\epsilon_{\alpha'}$. To facilitate the derivation, let us go into the interaction picture of $\hat{H}_0+\lambda\hat{H}_1$, denoted by the superscript $\mathrm{I}$, and write the Schr\"odinger equation as
\begin{align}
    i\partial_t\hat{U}_V^\mathrm{I}(t)=V\hat{H}^\mathrm{I}_G(t)\hat{U}_V^\mathrm{I}(t),
\end{align}
which is identical to an adiabatic evolution with the limit of large $V$ corresponding to the limit of large time. In the $V\to\infty$ limit, the relation $\hat{U}_{V\to\infty}^\mathrm{I}(t)\hat{P}^\mathrm{I}_\alpha(0)=\hat{P}^\mathrm{I}_\alpha(t)\hat{U}_{V\to\infty}^\mathrm{I}(t)$ is satisfied. In other words, $\hat{U}_{V\to\infty}^\mathrm{I}(t)$ maps $\mathcal{H}_{\hat{P}^\mathrm{I}_\alpha}(0)$ into $\mathcal{H}_{\hat{P}^\mathrm{I}_\alpha}(t)$, which in the Schr\"odinger picture formally reads
\begin{align}
    \ket{\phi(0)}\in\mathcal{H}_{\hat{P}_\alpha}\iff \ket{\phi(t)}\in\mathcal{H}_{\hat{P}_\alpha}.
\end{align}
As such, a state in a given sector at $t=0$ undergoes dynamics completely within that sector in the limit $V\to\infty$. Furthermore, one can employ the adiabatic theorem to derive that \cite{facchi2004unification-S}
\begin{align}
    \hat{U}_{V\to\infty}(t)=e^{-i[V\hat{H}_G+\sum_\alpha\hat{P}_\alpha(\hat{H}_0+\lambda\hat{H}_1)\hat{P}_\alpha]t},
\end{align}
up to a residual error of upper bound $\propto tV_0^2 L^2/V$, where $V_0$ is an energy term that is roughly a linear sum of $J$, $\lambda$, $\mu$, and $\kappa^2 S^2$ \cite{vandamme2021reliability-S}. From this, we arrive at the emergent gauge theory
\begin{align}\label{eq:H_QZE}
    \hat{H}_\mathrm{QZE}=\hat{H}_0+\lambda\sum_\alpha\hat{P}_\alpha\hat{H}_1\hat{P}_\alpha,
\end{align}
which will faithfully reproduce the dynamics under $\hat{H}=\hat{H}_0+\lambda\hat{H}_1+V\hat{H}_G$ up to a timescale $\propto V/V_0^2$, although we find in our exact diagonalization results that the timescale of this emergent gauge theory is $\propto V^2/(\lambda^2 J)$. This is not a contradiction since the prediction from the QZE formalism is a worst-case scenario. We note that in Eq.~\eqref{eq:H_QZE} we have further used the fact that $\sum_\alpha\hat{P}_\alpha\hat{H}_0\hat{P}_\alpha=\hat{H}_0$, since $\big[\hat{H}_G,\hat{H}_0\big]=0$, and we have left the term $V\hat{H}_G$ out of Eq.~\eqref{eq:H_QZE} as it trivially affects the dynamics of our observables.

It is instructive to elucidate the relation between the Zeno projectors $\hat{P}_\alpha$ and the gauge-invariant superselection-sector projectors $\hat{\mathcal{P}}_\mathbf{g}$. This relation is dictated by the choice of the sequence $c_j$. In principle, one may be able to find a sequence such that $\sum_jc_jg_j=\epsilon_\alpha$ is unique for every $\mathbf{g}$, in which case every $\hat{P}_\alpha$ corresponds to exactly one $\hat{\mathcal{P}}_\mathbf{g}=\hat{P}_\alpha$. However, such sequences would almost certainly have to be random or spatially inhomogeneous, and may involve energy scales that increase with system size to avoid resonances that may occur at higher orders. Nevertheless, as we have demonstrated numerically in the main text, in practice it can  suffice to use a simple sequence $c_j=(-1)^j$, which renders $\hat{H}_G$ fully translation-invariant. In such a case, a given $\hat{P}_\alpha$ will be the sum of all superselection-sector projectors $\hat{\mathcal{P}}_\mathbf{g}$ that satisfy $\sum_jc_jg_j=\epsilon_\alpha$. So long as only a few sectors $\mathbf{g}$ satisfy the latter equality for each $\alpha$, the single-body gauge protection will efficiently induce quantum Zeno dynamics. More importantly, if the sectors within a given Zeno subspace do not couple up to first order in $\lambda\hat{H}_1$, then, within the regime of validity of the QZE, this Zeno subspace is decomposed further into disconnected subspaces, namely the superselection sectors, and now the emergent gauge theory can be written as
\begin{align}\label{eq:H_QZE_final}
\hat{H}_\mathrm{QZE}=\hat{H}_0+\lambda\sum_\mathbf{g}\hat{\mathcal{P}}_\mathbf{g}\hat{H}_1\hat{\mathcal{P}}_\mathbf{g},
\end{align}
which is Eq.~\eqref{eq:QZE} that we introduced in the main text.

Let us now comment further on the physical picture arising from this quantum Zeno dynamics. In the $V\to\infty$ limit where quantum Zeno dynamics rigorously holds, the time-evolution operator becomes diagonal with respect to the gauge protection $V\hat{H}_G$. Consequently, an effective superselection rule arises, which, for an appropriate sequence $c_j$, is approximately the gauge-invariant superselection sectors due to $\hat{G}_j$. This superselection rule splits the Hilbert space into subspaces $\mathcal{H}_{\hat{P}_\alpha}\approx \mathcal{H}_{\hat{\mathcal{P}}_\mathbf{g}}$ that are invariant under the evolution. This suppresses gauge-breaking processes due to $\lambda \hat{H}_1$ indefinitely ($V\to\infty$ limit), thereby preserving disorder-free localization for all evolution times.

We note that in principle we cannot ascertain that our single-body gauge-protection scheme will work in the thermodynamic limit. However, we have two main reasons to believe that it will perform well for large-scale models. The first is that the quantum Zeno effect has weak dependence on system size, since by construction it dynamically isolates the symmetry sectors from each other. The second reason is that single-body gauge protection with the sequence $c_j=(-1)^j$ has been shown to work in practice using infinite matrix product state calculations, which compute many-body dynamics directly in the thermodynamic limit, for all accessible evolution times when starting in a given gauge-invariant sector in spin-$S$ $\mathrm{U}(1)$ quantum link models \cite{vandamme2021reliability-S}. We leave this open to future investigations.

\end{document}